# Inflation as a function of labor force change rate: cointegration test for the USA

I.O. Kitov[1], O.I. Kitov[2], S.A. Dolinskaya[1]

**Introduction**

Inflation forecasting plays an important role in modern monetary policy in developed countries. There is a trade-off between theoretical and practical aspects of the forecasting –. Any practical application has to be based on a sound theoretical background, and every theory must concentrate empirical facts and everyday practice in a few quantitative relations. In reality, monetary authorities more often use expert judgments instead of theoretical models. These judgements are often based on some concepts empirically proved to be wrong such as the conventional Phillips curve, "expectation augmented" Phillips curve or NAURU approach. As a matter of fact, there is no unique and comprehensive model in mainstream economics and econometrics, which is able to explain observations and predict inflation in developed countries.

A study of inflation as an economic variable driven solely by labor force change has been carried out by Kitov (2006a-c, 2007) for the USA, Japan, France, and Austria. This study has confirmed the existence of a linear relationship between inflation, unemployment and labor force. In the USA, the inflation-labor force relationship is characterized by a two-year time lag. Regression analysis (Kitov, 2006b) has shown a high level of correlation ($R^2$>0.9) between the variables and the lowermost, among available forecasting models, out-of-sample root-mean square forecasting error (RMSFE) at a two-year horizon, which corresponds to the observed lag. An important advantage of the model consists in its perfect parsimony - single predictor is used to fit observations in several countries during the last 25 to 45 years. Moreover, autoregressive properties, extensively used in many economic and econometric models of inflation, are not necessary because the relation between labor force change and inflation is strict. It does not allow any deviation between observed and predicted inflation when both variables are measured precisely.

As a consequence, the existence of a strict link between inflation and labor force change denies the concept of rational expectation thoroughly adopted in economics, i.e. future values of inflation are known to the extent some past values of labor force level are known. At the same time, the strict relation allows for a partial inflation control through labor force control and provides clear foundations for a sound economic policy.

---

[1] Institute for the Dynamics of the Geospheres, Russian Academy of Sciences
[2] Warwick University, UK

Both variables are non-stationary, however, implying a possibility for the regression results to be spurious ones despite the existence of a theoretical background (Kitov, 2006b, 2007) based on the fact that personal income distribution in the USA does not change with time when normalized to the total population and GDP. Therefore, econometric tests are necessary whether the relationship between inflation and labor force is a cointegrating one. The cointegration between inflation and labor force change would mean that the results of the previous regression analysis hold from the econometric point of view. For the USA, cointegration analysis is essentially a bivariate one. This significantly simplifies the implementation of relevant analysis and statistical inferences. However, in the case of modeling inflation using labor force change rate, a straightforward application of econometric methods potentially leads to wrong conclusions.

The rest of the paper is organized in four sections and one appendix. Section 1 briefly introduces the model and presents some results obtained for the USA in the previous study. Section 2 is devoted to the estimation of the order of integration in measured inflation and labor force change rate. Unit root tests are carried out for original series and their first differences. GDP deflator represents inflation in the study.

In Section 3, the existence of a cointegrating relation between the measured and predicted inflation is tested. The latter is represented by four time series reflecting improvements in the accuracy of prediction associated with suppression of random noise in the labor force readings. The presence of a unit root in the difference (residual) between the measured and predicted inflation implies the absence of cointegration between the variables and a strong bias in the results of the previous regression analysis. The residuals obtained from regressions of the measured inflation on the predicted ones are also tested for the unit root presence. This approach is in line with that proposed by Engle and Granger (1987).

The maximum likelihood estimation procedure developed by Johansen (1988) is used to test for the number of cointegrating relations in a vector-autoregressive representation (VAR). The existence of a cointegrating relation is studied as a function of the predictor

smoothness – from the original predictor to its three-year moving average. VAR and vector-error-correction models (VECM) are estimated for the forecasting purposes.

Section 4 discusses the principal results and their potential importance for economic theory and practical application, and concludes. Appendix introduces a pre-determined variable (from linear to a quasi-stochastic one) trend, which is of a crucial importance for cointegration study of time series related to population estimates such as labor force level.

1. **The static model**

There is a unique link between inflation and labor force change rate in the USA as obtained for the last 45 years (Kitov, 2006c). In econometric sense, it is a long-run equilibrium relation or static model which uses only original values of the involved parameters not their first differences. It is advisable to use the rate of labor force growth (not increment) as a predictor in order to match the dimension of inflation – $[t^{-1}]$. An implicit assumption of the model is that inflation does not depend directly on parameters associated with real economic activity. Moreover, inflation does not depend on its own previous and/or future values because it is completely controlled by a force of different nature – by the dynamics of the portion of population participating in labor force.

As defined by Kitov (2006a), inflation is a linear and potentially lagged function of labor force change rate:

$$\pi(t) = A_1 dLF(t-t_1)/LF(t-t_1) + A_2 \quad (1)$$

where $\pi(t)$ is the inflation at time $t$, $LF(t)$ is the labor force level at time $t$, $t_1$ is the time lag between the inflation and labor force, $A_1$ and $A_2$ are country-specific coefficients, which have to be determined empirically. The coefficients may vary through time for a given country as different units of measurements (or definitions) of the studied variables are used, but for true values of inflation and labor force they are constant.

Linear relationship (1) defines inflation separately from unemployment. This representation is an adequate one for the purpose of the paper. However, inflation and unemployment are two indivisible features of a unique process (Kitov, 2007). The process is the growth of labor force level, which is accommodated in real developed economies via two channels. The first channel is the increase in employment and corresponding change in personal income distribution (PID). All persons obtaining new paid jobs or their equivalents presumably change their incomes to some higher levels. There is an ultimate empirical fact, however, that the US PID practically does not change with time in relative terms, i.e. when normalized to the total population and total income (Kitov, 2005a,b). The increasing number of people at higher income levels leads to a certain disturbance in the PID. This over-concentration (or over-pressure in physical terms) of population in some income intervals above its neutral value must be compensated by such an extension in corresponding income scale, which returns the PID to its original density. Related stretching of the income scale is called inflation (Kitov, 2006a). The mechanism responsible for the income scale stretching, obviously, has some relaxation time, which effectively separates in time the source of inflation, i.e. the labor force change, and the reaction, i.e. the inflation itself.

The second channel is inherently related to those persons in the labor force who tried but failed to obtain new paid jobs. These people do not leave the labor force joining unemployment. Supposedly, they do not change the PID because they do not change their incomes. Therefore, the net labor force change equals the unemployment change plus the employment change, the latter process expressed in economies as the lagged inflation.

Kitov (2006c) proposed to estimate coefficients in (1) not by a linear regression of inflation on labor force change rate. Such a regression balances measurement errors, i.e. the deviations from a true relationship, through the entire period of observations (between 1960 and 2004) and has two important drawbacks: estimates of linear regression coefficients are usually biased due to errors in the predictor and the influence of larger fluctuations is overestimated. The latter might be of a higher importance in our case, considering the practice of revisions carried out by the US Bureau of Labor Statistics (BLS) and the Census Bureau (CB).

An alternative to regression arises from the fact that the labor force level is actually measured in monthly surveys not the change rate. Absolute uncertainty associated with annual measurements of the labor force level is constant or improving with time. Therefore, relative uncertainty associated with the net change in the labor force level during a given period is inversely proportional to the net change. When the net change in labor force level is large enough, as observed in the USA between 1960 and 2004, one obtains a significant improvement in the accuracy of the estimates of cumulative values. Considering measurement errors, a simple summation of annual estimates is always less accurate than a direct measurement of corresponding levels. (In the case of labor force, however, the revisions conducted by the BLS effectively cancel corresponding measurement errors out through time and provide practically the same accuracy of cumulative change for both the summation and the net level change techniques.) Therefore, the best-fit coefficients in (1) have to provide the same cumulative values for the measured and predicted inflation at the end of period, the latter having a decreasing uncertainty with time. This procedure should result in somewhat more accurate estimates, which are also different from those obtained by a standard linear regression. The estimated long-run relationship between inflation and labor force for the USA is as follows (Kitov, 2006c):

$$\pi(t)=4.0*dLF(t-2)/LF(t-2)-0.03075 \qquad (2)$$

The left-hand side of the equation is called below the *measured* inflation, $\pi_m(t)$, and the right-hand side – the *predicted* one, $\pi_p(t)$. A statistical assessment of the linear relationship carried out for the USA indicates that RMSFE of inflation forecasts at a two-year horizon for the period between 1965 and 2002 is only 0.8%. This value is superior to that obtained with any other inflation model, including the random walk approach, by almost a factor of 2. The entire period was also split into two segments before and after 1983. The forecasting superiority is retained with RMSFE of 1.0% for the first (1965-1983) and 0.5% for the second (1984-2002) sub-period.

There is one aspect not analyzed in the previous papers - both variables are non-stationary. Hence, validity of the regression results is under doubt (Granger & Newbold, 1967). When regressed, two I(1) processes may show unreasonably high correlation or goodness-of-fit and severely biased estimates of corresponding regression coefficients. This might happen to the studied relationship despite the sound theoretical foundation linking such macroeconomic variables as real GDP growth and inflation to population parameters only (Kitov, 2005ab, 2006ad). These links are treated as "mechanical" ones; i.e. any change in defining population parameters is always fully transmitted into a strictly proportional change in corresponding macroeconomic variables.

There are several predicted inflation time series obtained by Kitov (2006c) for GDP deflator and CPI measurements using corrected labor force readings. Corresponding corrections in labor force level include redistribution of step-like revisions to population controls carried out by the US CB and the BLS. These corrections smooth original labor force series (SA) and effectively removes the largest spikes from its first difference. Probably, the only exception is the years between 1990 and 1993, when an unprecedented number of revisions to population estimates and labor force survey methodology were carried out. The standard procedure of redistribution failed to iron the spikes out. So, the corrected labor force time series still contains an artificial spike near 1990. (The revision carried out by the BLS in July 2006 effectively resolved the problem of 1993, however.)

The set of the predicted time series also includes the one based on half-year shifted labor force measurements. This time series is better synchronized with inflation measurements because it averages monthly readings from July to June of the next year with December being in the middle of the period. The shifted series gives a larger $R^2$ than the original one presented by the BLS (Kitov, 2006c).

## 2. The unit root tests

It is clear from the behavior of the measured and predicted (in fact, labor force change rate) inflation curves displayed in Figure 1 that both time series are potentially characterized by the presence of unit roots because they demonstrate some clear non-stationary features. Moreover, the cumulative measured and predicted inflation curves are definitely non-stationary with a strong (stochastic or deterministic) trend component. In such a situation, a spurious regression is probable as modern econometric research shows. Therefore, some specific tests have to be carried out in order to prove validity of the results obtained in Kitov (2006b,c). In particular, one has to prove that the measured and predicted time series are integrated of order one, I(1), and are cointegrated, i.e. the residuals of their regression or VAR representation create a stationary, I(0), series which meets a number of specific requirements (Hendry&Juselius, 2001).

From statistical point of view, the principal problem of the econometric analysis related to the relationship between inflation and labor force change consists in the balance between the length of the series and the uncertainty of corresponding readings. Standard statistical tests work relatively well with time series of several hundred readings. When the size of a sample is small, statistical inferences based on asymptotic distributions provided with econometric packages may be biased. The replacement of asymptotic distributions by those obtained specifically for small samples reduces the power of the tests. Increasing the number of readings via sampling rate one observes decreasing accuracy. When measurement uncertainty is high no useful signal (i.e. long-term relationship) can be retrieved from the noise. There is no general solution of the problem and one always has to compare the outcome of statistical tests with the bulk of external information - theoretical and empirical.

As a first step in the analysis, we have to prove that the measured and predicted inflation are I(1) processes. There is a number of standard unit root tests provided with many econometric packages – the Dickey-Fuller, the Augmented DF (ADF), the modified DF t-test using a generalized least-squares regression (DF-GLS), Phillips-Perron and others. From this set, we use only the ADF and DF-GLS because the DF has a skewed distribution with a long left tail complicating discrimination the null of a unit root, and the PP test works better asymptotically, i.e. is not reliable at short time series. Standard programs of Stata9 statistical package are used in the current study.

There are several time series modeled in (Kitov, 2006c), which may be potentially tested for the unit root presence: the measured inflation (GDP deflator and CPI inflation) for the period between 1960 and 2004 (45 readings); numerous time series predicted from the labor force change rate ($dLF/LF$) – original, shifted by half a year, and those smoothed by moving average of various width. All the predicted time series are also present in two intervals: from 1958 to 2002 and from 1960 to 2004. The latter series are the versions of the former ones but shifted by two years ahead in order to synchronize the predicted inflation readings with corresponding measured values because of the two-year lag. The original predicted series retain the actual lead ahead of the measured values important for causality tests. Because the original and the two-year shifted time series differ by four values (each of the series contain two readings different from its counterpart) one may potentially expect a slightly different result of the unit root tests. From all available time series, measured and predicted, only those associated with the annual readings of GDP deflator are studied here.

Table 1 lists results of the ADF and DF-GLS tests for a unit root in the measured GDP deflator and six predicted inflation time series including *MA(2)* and *MA(3)* of the predicted time series which are shifted by two years ahead. The series *measured* represents the GDP deflator time series between 1960 and 2004, the series *predicted* and *shifted* are synchronized (shifted by two years ahead) with the *measured* one, and the original (not shifted) predicted series are titled as *predicted2* and *shifted2*. The last raw in the Table lists 1% critical values for corresponding tests. For the DF-GLS test, only values for lags 1 and 4 are presented because results for lags 2 and 3 are very close to those for lag 4.

One can derive an overall conclusion from Table 1 that no one series is a stationary one because the null of the unit root presence can not be rejected. It is worth noting that the predicted series shifted by two years, such as the *predicted* and *predicted2*, actually give different but close results. It does not affect the general conclusion of non-stationarity, however.

The series displayed in Table 1 are non-stationary. But it does not mean that these series are I(1) and some additional efforts are necessary to prove the assumption. For an I(1) process, the first difference has to be a I(0) process. Therefore, the same tests are repeated on the first differences of the two principal time series – *dmeasured* and *dpredicted,* which are displayed in Figure 2. Corresponding test results are presented in Table 2.

A standard unit root test usually contains many specifications related to various features potentially present in time series. The tests listed in Table 2 use two different assumptions on the presence of constant and trend in the first differences. First, we allow for a deterministic linear trend in the processes and test for a single unit root in an AR(p) representation, where p=0,1,2,3 for the ADF test and p=1,2,3 for the DF-GLS test. All the tests for the d*measured*, except those with lag 3, imply the rejection of the null of the unit root presence at the 1% level. For the *dpredicted*, only tests with lags 0 and 1 reject the null.

The presence of a constant term is a more reasonable hypothesis for the first differences of such processes as inflation and labor force change. When a constant specification is used in the ADF unit root tests, results are more favorable for the rejection of the null for both *dmeasured* and *dpredicted*. For the DF-GLS, the null is rejected for all lags in the *dmeasured* and only for lag 1 in the *dpredicted*. Because of the mixed results for the *dpredicted*, more information is necessary for rejection or acceptance of the null.

Here we have to draw a bold separation line between the procedures for estimation of GDP inflation and labor force change rate. Inflation is measured by acquisition of information on prices. There are some short-term corrections related to the inflation measurements which incorporate late information and new methodology. There is no correction extended far in the past, however, which is dependent on new inflation readings. Thus, one can consider subsequent readings of inflation as relatively independent in view of the measuring procedures.

This is not the case with the labor force measurements, however. The labor force level is estimated in monthly surveys using some limited population samples. Simultaneous estimates of age-gender-race distribution, then converted into population controls, are used for mapping the limited sample results onto the entire population. Such a construction of the labor force estimates has two drawbacks – the sample changes are slow in order to study the dynamics of the inter-sample evolution, and the population controls are determined both from the monthly-quarterly-yearly population estimates and decennial censuses.

The sample stability implies the presence of a link between subsequent labor force estimates. Corrections to the population controls are of even larger influence on labor force readings, however. After every decennial census, the BLS conducts severe (sometimes several per cent) revisions of the population age structure, which are extended long in the past. These revisions effectively include synchronized changes in the population controls over the entire period. Figure 3 demonstrates a remarkable effect of such revisions on the change in the number of people in a one-year-wide population cohort – the difference between 15-year-olds and 14-year-olds a year before. This age is of importance for the labor force measurements because statistics starts at 15 years of age. Therefore their influx is one of the defining components of the labor force increase along with net migration and total deaths. The difference between the numbers is everything but of a natural shape. Considering the disturbances introduced into the labor force time series by the BLS one should not be surprised that the DF-GLS tests does not reject the null for the first difference of the predicted inflation, which is based on the labor force change rate. Some statistical properties of population time series important for the cointegration analysis are discussed in Appendix.

The *dmeasured* time series is characterized by a very low average (0.00019±0.011) compared to the "min" and "max" values of -0.036 and +0.031. Slope and intercept of the regression on time are -0.00012±0.00012 and 0.24±0.24, respectively, i.e. the null of zero slope and intercept can not be rejected. This result implies that the *dmeasured* has no trend and intercept and the assumption of constant specification is applicable to the previous tests. The overall conclusion is that the *dmeasured* series has no unit root and the *dpredicted* series would not have a unit root if not the artificial distortion induced by the labor force measurement procedure. Therefore, the original series *measured* and *predicted* are integrated of order one.

### 3. The cointegration tests

The assumption that inflation and labor force change in the USA are two cointegrated non-stationary time series is equivalent to the assumption that the difference between the measured and predicted inflation, $\varepsilon(t)=\pi_m(t)-\pi_p(t)$, is a stationary or I(0) process. It is natural to start with a unit root test in the difference. If $\varepsilon(t)$ is a non-stationary variable having a unit root, the null of the existence of a cointegrating relation can be rejected. Such a test is associated with the Engle-Granger's approach, which requires $\pi_m(t)$ to be regressed on $\pi_p(t)$ as the first step, however. Since the predicted variable is obtained by a procedure similar to a linear regression and provides the best fit for cumulative values, we skip the regression and start with an analysis of the difference.

The hypothesis of a unit root is tested by the same procedures as before – the ADF and DF-GLS. If the null of a unit root is rejected, the hypothesis of no-cointegration is also rejected. In this case, the equilibrium relationship between the measured and predicted inflation is valid and a vector error-corrected model can be estimated using the results of the first stage.

Table 3 presents some results of the unit root tests. There are three differences tested: $diff1=(\pi_m(t)-\pi_p(t))$, $diff2=(\pi_m(t)-MA(2))$, and $diff3=(\pi_m(t)-MA(3))$. The ADF and DF-GLS results indicate the absence of a unit root in the *diff1* residual series, except may be for lag 3 in the DF-GLS tests. The *diff1* series of the largest importance because it is based on the inflation predicted from the original labor force series. As discussed in Section 2 and Appendix, the labor force measurements are potentially biased by a strong autocorrelation introduced by the revisions to the population controls. One can expect the enhancement of the autocorrelation in the differences (Chiarella&Gao, 2002). Moreover, this autocorrelation has to be more prominent in the smoothed series, where random noise in measurements is suppressed and piece-wise systematic deviations are of relatively higher amplitude. The series *diff2* and *diff3* illustrate these effects – the DF-GLS tests produce the results which are higher than corresponding 1% critical values also listed in the Table. So, smoothing leads to deterioration in the test results compared to those for the original series.

It is worth noting that the moving average time series provide a better description of the observed inflation as represented by standard deviation in the residual series. The best fit is obtained for the *MA(3)* series with standard deviation twice as low as that provided by the original predicted series: from 0.017 for *diff1* to 0.009 in *diff3*. The price paid for the better description is higher values in the unit root tests. Therefore, one can obtain a wrong statistical inference by over-suppressing of the stochastic component in real data, as shown in Appendix. Concluding the discussion, we reject the null of the unit root presence in the three difference series.

The next step is to use the Engle-Granger's approach and to study statistical properties of the residuals obtained from linear regressions of the measured inflation on various versions of the predicted inflation. Table 4 summarizes principal results of the regression analysis – coefficients with their standard deviations, $R^2$, RMS(F)E (F stays for forecasting because of the two year lag), and *t* for the null of zero coefficients. The dependent variable is always the *measured* series and predictor varies from the *predicted* to *MA(3)*. Results of specification tests on heteroskedasticity, omitted variables, ARCH effects, and serial correlation, also listed in Table 4, indicate that the residuals of the regressions, except those related to *MA(3)*, meet the requirements defining I(0) process and have properties of a white-noise realization necessary for a VAR representation of the measured and predicted inflation. Therefore, one cannot reject the null that the dependent variable and the predictors are cointegrated. The reasons for the failure of the specification test for *MA(3)* have been already discussed.

The Johansen's (1988) approach is based on the maximum likelihood estimation procedure and tests for the number of cointegrating relations in the vector-autoregressive representation. The above analysis has shown that the VAR representation is an adequate one due to the properties of the regression residuals. The Johansen's approach allows simultaneous testing the existence of cointegrating relations and determining their number (rank). For two variables, only single cointegrating relation is possible. When cointegration rank is 0, any linear combination of the two variables is a non-stationary process. If rank is 2, both variables have to be stationary. Thus, when the Johansen test results in rank 1, a cointegrating relation does exist.

Table 5 lists trace statistics, eigenvalues, LL and other information obtained from the cointegration rank tests for a number of predictors and trend specifications. The maximum number of lags included in the underlying VAR model is 4 in all tests. No null of cointegration rank 1 can be rejected. So, there is a cointegrating relation between the measured and predicted inflation.

Because of the two-year lag behind the labor force change and the existence of the cointegrating relation or long-run equilibrium relationship (2), a test on causality direction is a trivial one. In this test, the variables *predicted* and *shifted* are replaced with their original version *predicted2* and *shifted2*, which leads the *measured* variable by two years. Results of the causality tests are presented in Table 6, which demonstrates that the predictors are weakly exogenous variables.

Now we are sure that the measured and predicted inflation series are cointegrated and the principal results of the previous research hold. Therefore, the estimates of the goodness-of-fit and RMSFE are accurate. The estimates were obtained in a simplified regression procedure, which does not use autoregressive properties of noise. A standard moving average smoothing provides a substantial suppression of the noise, as the increase in $R^2$ from 0.7 for the annual readings to 0.9 for *MA(3)* demonstrates (Kitov, 2006c).

A VAR representation may potentially provide a further improvement due to additional noise suppression. In practice, AR is a version of a weighted moving average, which optimizes noise suppression throughout the whole series. In the VAR model, we use the *shifted2* series as an exogenous predictor and the maximum lag 4 instead of the pre-estimated one maximum lag 3. The level of correlation between the measured and predicted series does not allow any past values of inflation to influence the present ones and, hence, the past values of labor force change. The equilibrium relationship is a strict one and converges to a unique curve when measurement errors are eliminated. At any time, inflation does not contain information important for labor force change. So, the labor force change rate is an exogenous variable. One cannot deny the influence of current inflation on current labor force level. This influence is very weak, however, as this study shows.

The distribution of the VAR error term is close to the normal distribution with skewness=-0.05 and kurtosis=3.99, the former being of much higher importance for the Jarque-Bera normality test and validity of statistical inference. The VAR stability is guaranteed by the eigenvalues of the companion matrix lower than 0.54. There is practically no autocorrelation at lags from 1 to 4 as follows from the LM statistics. Therefore the VAR model accurately describes the data and satisfies principal assumptions on the residuals.

A VECM representation uses additional information to that provided by the VAR models due to separation of noise and equilibrium relationship. So, it potentially provides an improvement on the VAR models. Tables 7 and 8 list some results of the VAR and VECM for a number of predictors. There is only a marginal improvement in RMSFE on that obtained by the linear regressions. The smallest RMSFE is 0.0076 in the VAR representation (*MA(3)* as a predictor) and 0.0073 in the VECM representation (*shifted2* as a predictor). The best RMSFE from the linear regressions for the period between 1965 and 2002 is 0.008.

One can conclude that these powerful statistical methods fail to improve on the simple predictions. A significant decrease in the forecasting uncertainty is possible only through a major increase in the accuracy of inflation and labor force measurements. This conclusion is in line with a standard research cycle in hard sciences.

## 4. Conclusion

The core result of the analysis consists in formal statistical confirmation of the existence of a unique linear and lagged (two years for the USA) relationship between inflation and labor force change. Hence, the two I(1) variables are cointegrated in statistical sense; i.e. their residual time series is proved to be a stationary one with white noise characteristics - no unit root, normal underlying distribution, no omitted variables, no autocorrelation, no heteroskedasticity. The absence of such cointegration test was a weak point in the previous papers. It exposed the results of modeling and theoretical consideration to criticism.

Observed inflation is always a combination of two components – a pre-determined variable trend (demonstrating stochastic features), which one-to-one repeats the behavior of some true labor force change, and a noise component related to measurement errors in both inflation and labor force level. These errors include random and piece-wise systematic parts.

In terms of physics, there is a process "mechanically" linking true inflation to true labor force change. Observed deviations from the relationship describing the mechanical link are measurement errors. The long-run equilibrium relationship implies the absence of any structural breaks necessary for the explanation of the changes in the US inflation observed during the last 50 years. Such structural breaks are a common feature of a majority of models developed in the econometric framework, where inflation is considered as a stochastic process.

The noise component demonstrates some artificial features related to the procedures used by the US BLS for the labor force estimation. As in hard sciences, improvements in the measurement procedures, both for inflation and labor force, can potentially provide any desired accuracy of inflation forecast at a two-year horizon and even further if to improve the accuracy of labor force projections. It is worth noting that there is no error-correction mechanism associated with the deviations of true inflation values from the true predicted ones, but there exist error-correction mechanisms, which compensate random and systematic errors through time. For the labor force series, this mechanism is obviously associated with multiple revisions, which reshape the error pattern, in amplitude and timing, at very long time horizons.

The long-run relationship between inflation and labor force change rate is explained by the reaction of personal income distribution on the disturbance induced by those persons who enter employment and change their incomes correspondingly. This consideration is a part of the new economic concept, which relates macroeconomic variables to quantitative characteristics of population only.

**Appendix. Variable deterministic trend**

In hard sciences, situations where a non-stationary time series practically repeats the behavior of another series with some time lag are common. Finding of such a resemblance is a first step of a variety of research programs. The next natural step is to reduce the level of noise associated with measurement errors. For a scientist, this step is necessary for obtaining a higher confidence in the existence of a strict link between two variables or in the absence of such a link. So, improving measurement accuracy results in the acceptance of a true relationship or in the rejections of a false one.

In econometrics, however, the presence of such a pre-determined variable trend often leads to deterioration of the performance of standard statistical procedures associated with non-stationary processes. The danger to wrongly reject the presence of a true link in standard econometric procedures is associated with a general rejection of the existence of real deterministic trends in economic time series except the simple ones as connected to economic growth. Such simple trends are easily removed by detrending procedures. Otherwise, there exist only stochastic trends in econometrics, which represent the central problem in analysis of non-stationary time series. In some cases, this assumption leads to a biased result. For example, when correlation between two series in strong and random measurement errors are significantly suppressed, even small systematic errors induced by estimation procedures become defining for rejection of cointegration, i.e. for rejection of the true link between the variables.

There are several key assumptions, which lay in the basement of the cointegration concept. One of them is associated with the properties of residuals necessary for two or more non-stationary processes to be cointegrated. The existence of a cointegrating relation means that the residual of some linear combination of non-stationary variables is stationary. This residual time series has to be an independent and identically distributed with zero mean and constant variance, $IID(0,\Omega^2)$, where IID is not necessary the normal distribution. Such residuals guarantee unbiased estimates of relevant coefficients.

The mainstream research of cointegration phenomenon allows for some deterministic trend in non-stationary variables which is very often considered as a linear or quadratic one. The inclusion of such trend terms sometimes helps to distinguish between purely stochastic non-stationary processes and stationary processes with deterministic trends. There are some studies of more complicated cases with nonlinear deterministic trends. But most of them consider analytic functions. Here we extend the notion of deterministic trend to the level of deterministic variable (potentially stochastic) process, i.e. to such a process, which has stochastic statistical properties but is fully pre-determined. In physics, this notion is similar to decomposition of a measured time series into true and noise components.

The importance of deterministic variable trend for econometrics can be illustrated by a simple but principal example associated with the discussion in Section 2 of growth in the number of 15-year–olds in the USA, $N15(t)$. The evolution of this number is definitely described by a stochastic time series, supposedly, by I(2) process. In the framework of econometrics, this implies that the expected value of the change in $N15$ is zero (if to exclude a small deterministic trend):

$$N15(t)-N15(t-1)=dN15(t)=\varepsilon(t) \qquad (A1)$$

where $\varepsilon(t)$ some stochastic ID$(0,\sigma^2)$ process. (Actual $dN15$ average value between 1960 and 2004 is 29900 and standard deviation is 193000. If to exclude the years between 1960 and 1962, the mean=14500 and stdev=115000.) Thus, $dN15(t)$ must have very low level of autocorrelation for the small time series under consideration. Asymptotically, autocorrelation at any lag must be zero – $E[(\varepsilon(t)*\varepsilon(t-s)]=0$ for any s≠0. In fact, autocorrelation associated with the actual series of $dN15(t)$ between 1960 and 2004 is below 0.25 at lags from 1 to 15. Therefore, the annual increment in the $N15(t)$ can hardly be predicted from its previous values. This is a standard situation in econometrics, which deals mainly with economic and financial variables. A general convention dictates that there is no way to exactly predict the forthcoming change in stock prices or CPI inflation. Only statistical estimates are available. This is not the case with the $dN15(t)$, however.

First, one has to learn some statistical properties of the $N15$. Figure A1 shows the annual $N15$ readings, the first difference, $dN15$, and also the second differential $d2N15$. (There are also $N0$ and $N14$ and their differentials shown in the Figure.) Due to strong variations in the number of 15-year-olds in the earlier 1960s we limit the $N15$ series to the period between 1963 and 2004 which contains 42 readings.

The ADF test rejects the null of a unit root in the *d2N15* time series and accepts the null for *N15* (for lag 0 and larger) and *dN15* (for lag 1 and larger). The DF-GLS tests for the same series assume the acceptance of the null in the *d2N15*, however (see Table A1). Thus, the problem of a unit root in the *d2N15* is a controversial one and needs more attention to the population estimation process. For our purposes, we reject non-stationary in the *d2N15*.

Regardless of the unit root presence in the *N15* and its stochastic properties, one can accurately predict its value at a one (and more) year horizon using the change in the number of 14-year-olds one year ago, *dN14(t-1)=N14(t-1)-N14(t-2)*, as a proxy to the *dN15(t)=dN14(t-1)+ω(t)*, where *ω(t)* is associated with the differences in death and migration rate between 14-year-olds and 15-year-olds in the given year. Another way to obtain an accurate estimate of the *N15(t)* is to update *N14(t-1)* by the inflation-deflation method used by the Census Bureau in the population estimates. Actually, the *N15(t)* contains the same population cohort, by the year of birth, as the *N14(t-1)* plus net migration, *m(t)*, and less total deaths, *ρ(t)*, in the population of the given age. Therefore, one can rewrite relationship (A1) in the following forms:

$$N15(t)=N15(t-1)+dN14(t-1)+\omega(t) \qquad (A2)$$
$$N15(t)=N14(t-1)+\rho(t)+m(t) \qquad (A3)$$

Relative importance of *ω(t)* can be estimated from the difference between *dN15(t)* and d*N14(t-1)* normalized to the *dN14(t-1)*, as presented in Figure A2. The shape of the curve is not a surprise because of multiple revisions made by the USA Census Bureau. The years between decennial censuses are characterized by low values (<0.1) of the ratio *(dN15(t)-dN14(t-1))/dN14(t-1)* indicating that the fraction of the *dN15(t)* not explained by the *dN14(t-1)* is very small. Therefore, the net effect of *ω(t)* is below 10% for those 7 to 8 years when the disturbance induced by population revisions after decennial censuses is low.

The years around the censuses demonstrate a highly volatile behavior of the ratio. Such a behavior has a clear explanation – new counts obtained in decennial censuses have to be accommodated into new population estimates. As a rule, old (so-called postcensal) population estimates show age structures different from those obtained in the censuses for the years of censuses. A banal way to introduce the new structure is to make a break in the single year of age populations. Unfortunately for our approach, the breaks are made simultaneously in all age groups. As a result, the number of 14-year-olds one year ago is not corrected in the same way as the number of 15-year olds in the given year and their difference demonstrates a spike. One can eliminate such spikes by introducing an appropriate correction into the *N14(t-1)*, which can be effectively the same as the correction in the *N15(t)*.

When the spikes near the census years are removed, the difference *N15(t)-N14(t-1)* (and hence, the difference *dN15(t)-dN14(t-1)*) still looks biased as Figure A3 demonstrates. There are three intervals with constant differences: from 1981 to 1990, from 1991 to 2000, and after 2000, and one interval with a positive linear trend - between 1963 and 1980. Such a behavior apparently results from the methodology used by the Census Bureau to balance the total population growth over the whole age structure. It is difficult to believe, however, that the difference evolves in such a deterministic way. On the other hand, this deterministic behavior can be eliminated from the difference by subtracting the mean values in the periods of constant difference and by compensating the trend between 1963 and 1980. Figure A3 displays the corrected difference which now fluctuates around the zero line and has no trend. The mean value of the corrected difference *N15(t)-N14(t-1)* for the period between 1963 and 2004 is zero and standard deviation is only 2200 or 1.5% of that of that in *dN15*.

Thus, we found that a number of reasonable corrections can provide an almost precise estimate of the difference between the *N14(t-1)* and *N15(t)*. The two terms in (A3) associated with deaths and migration are decomposed in a deterministic and stochastic component, the former being effectively attached to the deterministic trend term *N14(t-1)*. Theoretically, one can consider this combination as a pre-determined variable trend - it demonstrates stochastic properties but is completely known in the given year *t*. In practice, one can use the current trend in the difference *N15(t)-N14(t-1)* (Figure A3) and current estimates of the *N14(t)* for a prediction of the *N15(t+1), t=2007,…,2010*.

The same statement is valid for the first differences of the *N15(t)* and *N14(t-1)*, where the latter is the pre-determined variable trend of the former. A very conservative amplitude estimate of the residual stochastic component of the difference *u(t)=dN15(t)-dN14(t-1)* would be 0.05*dN14(t-1)*, i.e. *u(t)=o(dN14(t))* for our purposes (see Figure A4). Such a small and random variation in the difference between *dN14(t-1)* and *dN15(t)* should give a very high level of correlation between the variables. A linear regression gives $R^2$=0.98, the slope of 0.996±0.003 and intercept 18±471.

The number of 15-year-olds is of a special importance in economics. This number represents a principal component of the annual increase in labor force along with effects of participation rate, net migration and deaths. Therefore, one can expect the labor force level demonstrating the existence of a nonzero deterministic variable trend. In addition to the pre-determined, according to (A2) or (A3), influence of the *N15(t)*, the labor force participation rate is characterized by a high degree of predictability, at least during the last 45 years. Figure A5 shows a period of practically linear growth between the mid-1960s and the mid-1990s induced by the active growth in the women's participation rate. In fact, a linear trend shown in the Figure explains more than 97% of the variability in participation rate during the period. The last five years are characterized by a slight downward trend in the participation rate. The USA BLS, the CBO and a number of other agencies and institutions provide a wide range of projections at various time horizons associated with the participation rate and labor force itself.

It is time to recall that the *dN15(t)* and *dN14(t-1)* series are I(1). For many economic and financial non-stationary time series, a regression of two I(1) processes may be spurious, i.e. to give severely biased estimates of regression coefficients and $R^2$. In our case, there is no doubt that the variables are practically identical and the estimates of the linear regression coefficients and goodness-of-fit are consistent. This implies that the variables have to be cointegrated, i.e. the difference between them has to be a I(0) process.

The estimation of cointegration rank in the framework of the Johansen's VAR methodology shows, however, that there is no cointegrating relation between the variables. Table 2A lists results of the cointegrating rank tests for the original and corrected series *dN15(t)* and *dN14(t-1)*. The tests reject the existence of any cointegration, i.e. cointegration rank 1, in both cases. This outcome, while surprising, obviously belongs to the type I spurious regression (Chiarella&Gao, 2002), i.e. rejection of a true relation. Such wrong rejection is driven by the artificial statistical properties of the difference between the variables induced by the Census Bureau revisions – original and those left after the corrections. It is worth noting that the original variables *N15(t)* and *N14(t-1)* also are not cointegrated, but the corrected series are integrated. It implies that two I(2) processes, as follows from the unit root tests of their first differences, have a I(0) difference.

Such situations deserve a special consideration in the framework of econometrics. Before starting any statistical study one has to be absolutely sure that the residual or errors are of significant amplitude. In some cases, residuals characterized by high autocorrelation and other possible statistically "wrong" properties should not prevent acceptance of strong correlation between economic variables. This is especially important for the variables associated with population surveys, where means are used in order to enhance these "wrong" statistical properties. Labor force estimates are not an exclusion from the list.

As a synthetic example of such a behavior, several time series are constructed mimicing principal features of the population and labor force estimates. A reference time series is described by the following function: *r(t)=0.3sin(0.1t)*, where *t*=1,…,45. Sinus function provides a degree of non-stationarity, when tested for the unit root presence, and has a period roughly corresponding to that of actual population time series: ~30 years. A piece-wise systematic error term of various amplitude but the same time structure is added to the reference function. Time dependence of the error term is described by constants of 10-year length with changing sign: $e(A,t)=A*(-1)^{int(0.1*t+1)}$, where *A* is coefficient varied from 0.005 to 0.1 in the study, *int(.)* is the integer part of the value in the brackets. The 10-year intervals correspond to decennial censuses.

So, we test for cointegration between the following time series: *r(t)* and *r(t)+e(A,t)* as a functions of *A*. First, the reference series is regressed on the disturbed series. As expected, goodness-of-fit is excellent: $R^2$ from 0.999 for *A*=0.005 to 0.784 for *A*=0.1. There is no doubt that the regression is not a spurious one in physical terms – the curves practically coincide. There is an obvious problem with statistical tests. One can expect that the residuals of the regression demonstrate strong, and independent on *A*, autocorrelation prohibiting the existence of a cointegrating relation between the studied functions, before A is large enough to avoid co-linearity threshold predefined in every statistical package. In fact, the Durbin-Watson test gives an almost constant value around 0.36 and LM tests for autoregressive conditional heteroskedasticity consistently rejects the null of no ARCH effects.Obviously, when a white noise component of practically any amplitude is added to *r(t)*, the same statistical tests give excellent results regardless the goodness-of-fit because the regression residuals now have properties of white noise.

This consideration is a trivial one from the point of view of econometrics and would not deserve any further efforts to be wasted. In reality, one often is not aware of specific properties of actual time series and still has to rely on statistical tests. Actual measurement noise very probably contains an artificial component, such as the periods of a constant difference observed in the population estimates, and a random component. Supposedly, relative amplitude of the latter is decreasing due to improving procedures. The random component can be also suppressed by smoothing. Therefore, the artificial component becomes defining for statistical estimates at some point. This definitely happens, when moving average is applied to smooth the labor force estimates. Goodness-of-fit increases, RMSE decreases, the predicted curve converges to the measured one, and … statistical tests start to reject cointegration.

**Tables**

Table 1. Unit root tests of the measured (GDP deflator) and predicted inflation in the USA

| Variable | ADF [lag 0] | DF GLS [lags] [1] | [4] |
|---|---|---|---|
| *measured* | -1.57 | -1.93 | -1.46 |
| *predicted* | -3.53 | -1.95 | -1.22 |
| *shifted* | -1.95 | -2.10 | -1.48 |
| *predicted2* | -3.46 | -2.14 | -1.55 |
| *shifted2* | -1.73 | -2.19 | -1.52 |
| *MA(2)* | -1.50 | -2.18 | -1.32 |
| *MA(3)* | -1.40 | -1.65 | -1.45 |
|  |  |  |  |
| 1% critical | -3.62 | -3.77 | -3.77 |

Table 2. Unit root test of the first differences of the measured and predicted inflation

|  | ADF [lag] | | | | DF-GLS [lag] | | |
|---|---|---|---|---|---|---|---|
| Variable | 0 | 1 | 2 | 3 | 1 | 2 | 3 |
| *dmeasured* (trend) | -5.02 | -5.62 | -4.25 | -3.56 | -5.51 | -4.16 | -3.44 |
| *dpredicted* (trend) | -5.89 | -6.87 | -4.04 | -3.62 | -4.83 | -2.53 | -2.39 |
| 1% critical | -4.20 | -4.21 | -4.22 | -4.23 | -3.77 | -3.77 | -3.77 |
| *dmeasured* (cons) | -5.18 | -5.44 | -4.06 | -3.25 | -5.32 | -3.96 | -3.22 |
| *dpredicted* (cons) | -5.78 | -6.70 | -3.86 | -3.45 | -3.00 | -1.47 | -1.33 |
| 1% critical | -3.62 | -3.63 | -3.64 | -3.64 | -2.63 | -2.63 | -2.63 |

Table 3. Unit root test of the differences between the measured and predicted inflation

| Variable | ADF | DF GLS [lags] | | | Mean | stdev | Pr(skew) | Pr(kurt) | chi$^2$ | Pr>chi$^2$ |
| --- | --- | --- | --- | --- | --- | --- | --- | --- | --- | --- |
| | | [1] | [2] | [3] | | | | | | |
| *diff1* | -7.60 | -4.89 | -2.88 | -2.11 | 0.00095 | 0.017 | 0.36 | 0.18 | 2.76 | 0.25 |
| *diff2* | -5.68 | -5.45 | -2.39 | -1.96 | 0.00088 | 0.012 | 0.39 | 0.46 | 1.34 | 0.51 |
| *diff3* | -4.09 | -4.01 | -2.41 | -2.44 | 0.00081 | 0.009 | 0.26 | 0.84 | 1.36 | 0.51 |
| 1% critical | -3.62 | -2.63 | -2.63 | -2.63 | | | | | | |

Table 4. Results of tests for heteroskedasticity, omitted variables, ARCH effects, and serial correlation as applied to the residuals of linear regressions of variable *measured* on four predictors.

| Predictor | Hettest [1) Pr>chi2 | Ramsey [2) test Pr>F | LM for ARCH [3) Pr>chi2 | Breusch-Godfrey LM [4) Pr>chi2 | DW [5) d-stat | $R^2$ | RMS(F)E | Cons [cons] Pr>\|t\| |
|---|---|---|---|---|---|---|---|---|
| *Predicted* | 0.21 | 0.0046 | 0.80 | 0.13 | 1.61 | 0.62 | 0.015 | 0.012 [0.004] 0.004 |
| *Shifted* | 0.25 | 0.0029 | 0.97 | 0.18 | 1.58 | 0.79 (0.94) | 0.012 | 0.005 [0.003] 0.15 |
| *MA(2)* | 0.13 | 0.043 | 0.37 | 0.12 | 1.55 | 0.83 (0.95) | 0.010 | 0.0014 [0.003] 0.66 |
| *MA(3)* | 0.29 | 0.0065 | 0.88 | 0.008 | 1.16 | 0.86 (0.96) | 0.0095 | 0.0008 [0.003] 0.78 |

1) $H_0$ - constant variance; 2) $H_0$ - no omitted variables; 3) $H_0$ - no ARCH effect; 4) $H_0$ - no serial correlation; 5) $H_0$ - no serial correlation

Table 5. Cointegrating rank of a VECM

| predictor | trend specification | rank | parms | LL | eigenvalue | trace statistics | 5% critical value |
|---|---|---|---|---|---|---|---|
| predicted | constant | 0 | 14 | 242.5 | . | 25.77 | 15.41 |
| predicted | constant | 1 | 17 | 254.0 | 0.427 | 2.9037* | 3.76 |
| shifted | constant | 0 | 14 | 265.3 | . | 21.08 | 15.41 |
| shifted | constant | 1 | 17 | 274.6 | 0.365 | 2.4773* | 3.76 |
| predicted | rconstant | 0 | 12 | 242.5 | . | 25.81 | 19.96 |
| predicted | rconstant | 1 | 16 | 253.9 | 0.428 | 2.9282* | 9.42 |
| shifted | rconstant | 0 | 12 | 265.2 | . | 21.19 | 19.96 |
| shifted | rconstant | 1 | 16 | 274.6 | 0.366 | 2.4830* | 9.42 |
| predicted | none | 0 | 12 | 242.5 | . | 22.68 | 12.53 |
| predicted | none | 1 | 15 | 253.6 | 0.419 | 0.4437* | 3.84 |
| shifted | none | 0 | 12 | 265.2 | . | 18.29 | 12.53 |
| shifted | none | 1 | 15 | 274.0 | 0.349 | 0.6744* | 3.84 |

Table 6. Granger causality test

| Variable | Excluded | chi$^2$ | Pr>chi$^2$ |
|---|---|---|---|
| measured | predicted2 | 15.2 | 0.000 |
| predicted2 | measured | 2.7 | 0.290 |
| measured | shifted2 | 37 | 0.000 |
| shifted2 | measured | 1.5 | 0.342 |

Table 7. VAR results for the measured GDP deflator.

| measured | predicted | | shifted | | MA(2) | | MA(3) | |
|---|---|---|---|---|---|---|---|---|
| | Coef. | Std. Err. | Coef. | Std. Err. | Coef. | Std. Err. | Coef. | Std. Err. |
| L1 | 1.015 | 0.154 | 0.743 | 0.174 | 0.699 | 0.174 | 0.545 | 0.140 |
| L2 | -0.313 | 0.139 | -0.205 | 0.130 | -0.315 | 0.126 | -0.129 | 0.106 |
| Exogenous [1] | 0.184 | 0.090 | 0.400 | 0.116 | 0.567 | 0.151 | 0.602 | 0.094 |
| cons | 0.005 | 0.003 | 0.003 | 0.003 | 0.002 | 0.003 | 0.000 | 0.002 |
| $R^2$ | 0.84 (0.95) | | 0.86 (0.96) | | 0.87 (0.96) | | 0.90 (0.98) | |
| RMS(F)E [2] | 0.01 (0.010) | | 0.0095 (0.0094) | | 0.0093 (0.0092) | | 0.0076 (0.0076) | |

1) Variables from *predicted* to *MA(3)* are weakly exogenous. Only two lags are used in every case.
2) RMS(F)E is the root mean square of error of description (in-sample error), which is obviously the forecasting error (out-of-sample error) at the same time due to the two-year lag of the inflation behind labor force change.

Table 8. VECM results for the measured GDP deflator

| Predictor | Coeff. | $R^2$ (measured) | $R^2$ (predictor) | RMS(F)E (measured) | RMS(F)E (predictor) |
|---|---|---|---|---|---|
| *predicted* [lag 2] | -1.04 [0.05] | 0.15 | 0.61 | 0.0110 | 0.0160 |
| *shifted* [lag2] | -1.07 [0.05] | 0.21 | 0.44 | 0.0103 | 0.0110 |
| *predicted2* [lag 4] | -1.07 [0.06] | 0.59 | 0.4 | 0.0080 | 0.0210 |
| *shifted2* [lag4] | -1.08 [0.06] | 0.66 | 0.39 | 0.0073 | 0.0110 |
| *MA(2)* [lag 2] | -1.09 [0.04] | 0.22 | 0.53 | 0.0100 | 0.0070 |
| *MA(3)* [lag 2] | -1.05 [0.05] | 0.43 | 0.10 | 0.0086 | 0.0090 |

Table A1. Results of the ADF and DF-GLS tests for *N15*, *dN15*, and *d2N15*

| Predictor | ADF [lag] (constant) | | | | | DF-GLS [lag] (with trend) | | | |
|---|---|---|---|---|---|---|---|---|---|
| | 0 | 1 | 2 | 3 | 4 | 1 | 2 | 3 | 4 |
| *N15* | -1.27 | -1.67 | | | | -1.38 | -1.73 | -1.93 | -1.68 |
| *dN15* | -5.18 | -3.15 | -2.73 | | | -2.89 | -2.47 | -2.54 | -2.14 |
| *d2N15* | -16.16 | -7.06 | -5.62 | -4.37 | -4.32 | -2.04 | -1.98 | -2.03 | -1.93 |
| 1% critical | -3.64 | -3.65 | -3.66 | -3.66 | -3.67 | -3.77 | -3.77 | -3.77 | -3.77 |

Table 2A. Results of cointegration rank test.

| Specification | [cons] | | [trend] | | [none] | |
|---|---|---|---|---|---|---|
| rank | 0 | 1 | 0 | 1 | 0 | 1 |
| *dN15(t)* vs. *dN14(t-1)* original | 36.62 | 11.61 | 39.05 | 11.61 | 35.95 | 11.22 |
| *dN15(t)* vs. *dN14(t-1)* corrected | 48.29 | 9.49 | 48.39 | 9.47 | 47.96 | 9.16 |
| *N15(t)* vs. *N14(t-1)* original | 11.60* | 2.62 | 12.58* | 1.65 | 7.61* | 0.16 |
| *N15(t)* vs. *N14(t-1)* corrected | 26.72 | 2.75* | 26.83 | 2.85* | 23.71 | 0.21* |
| 5% critical | 15.41 | 3.76 | 18.17 | 3.74 | 12.53 | 3.84 |

**Figures**

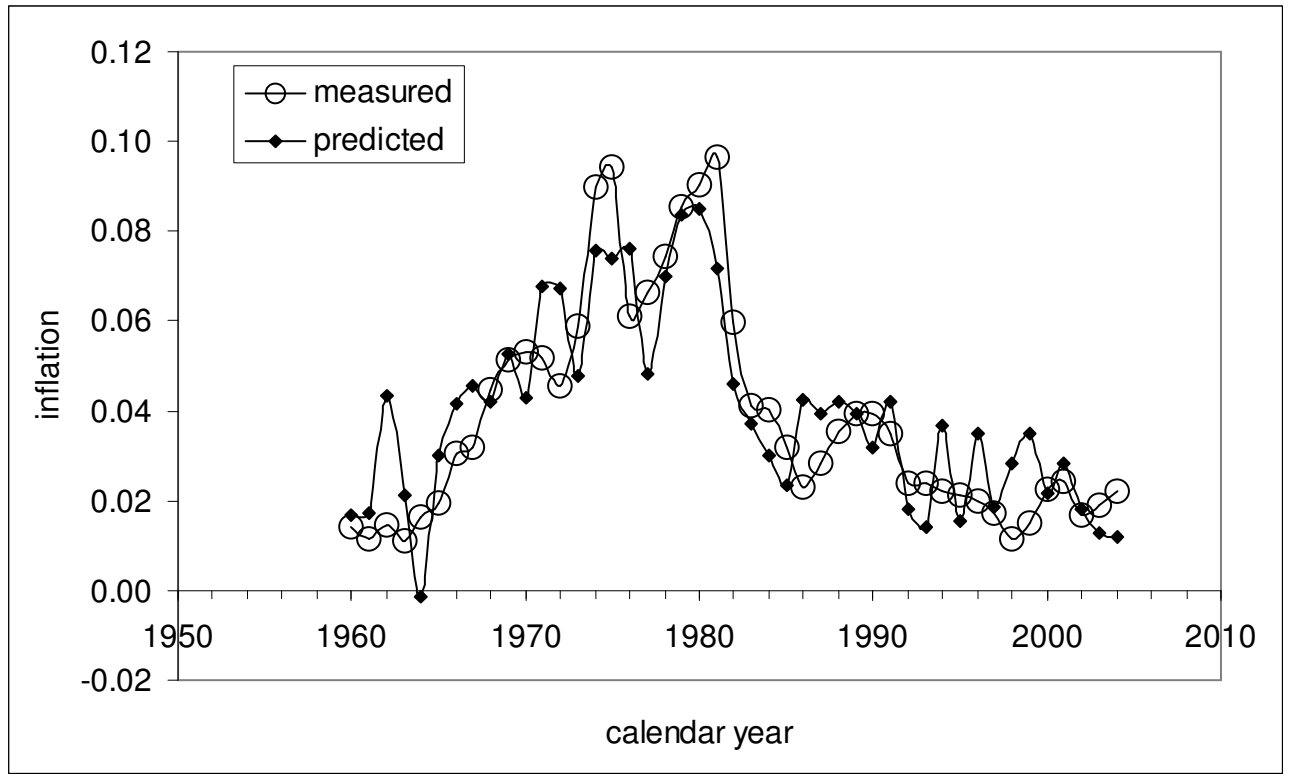

Figure 1. The *measured* and *predicted* time series between 1960 and 2004. The labor force estimates are obtained at the BLS web-site in July 2006. Therefore the strong spike near 1993 discussed in (Kitov, 2006c) has disappeared.

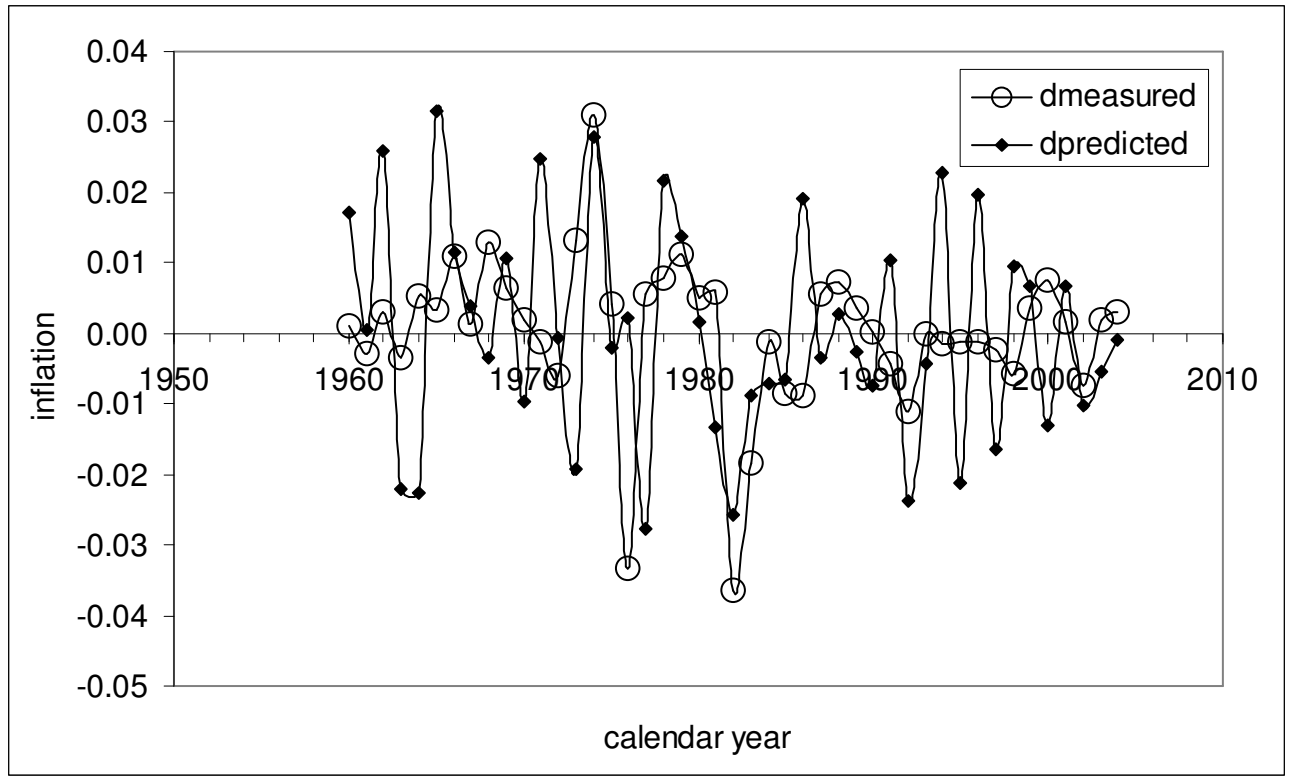

Figure 2. The series *dmeasured* and *dpredicted* obtained as the first differences of the *measured* and *predicted* in Figure 1.

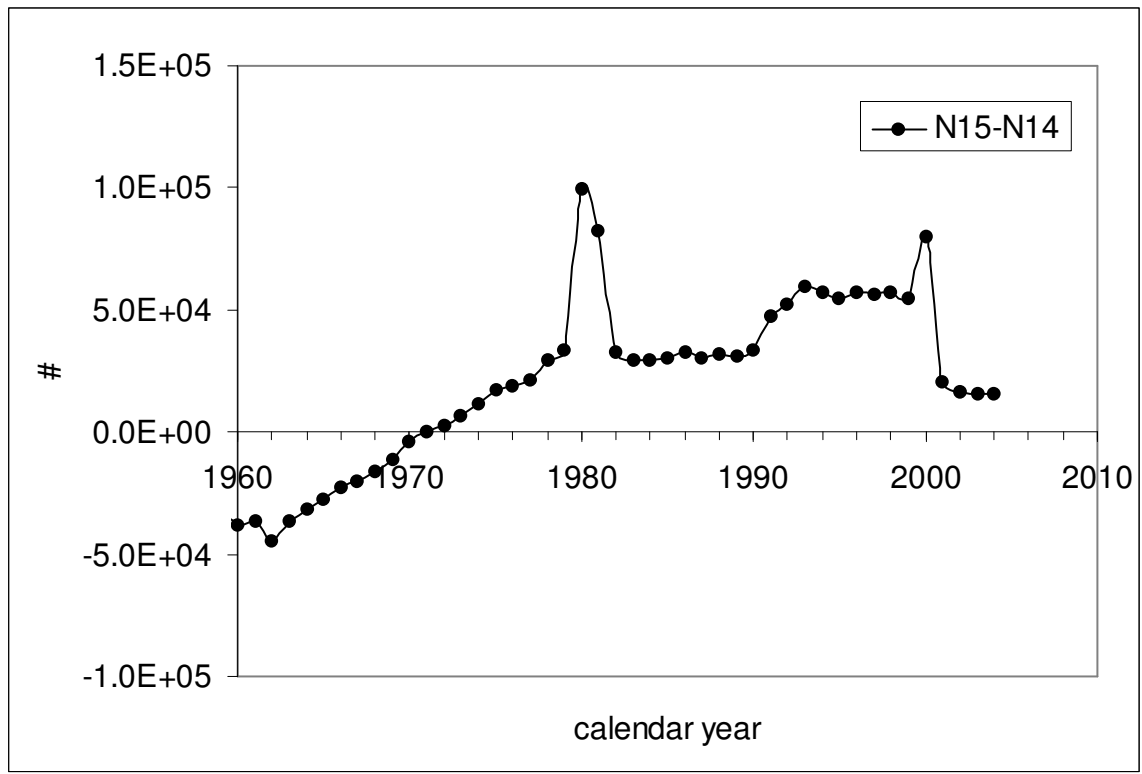

Figure 3. The difference between *N15(t)* and *N14(t-1)* in the USA for the period between 1960 and 2004

a)

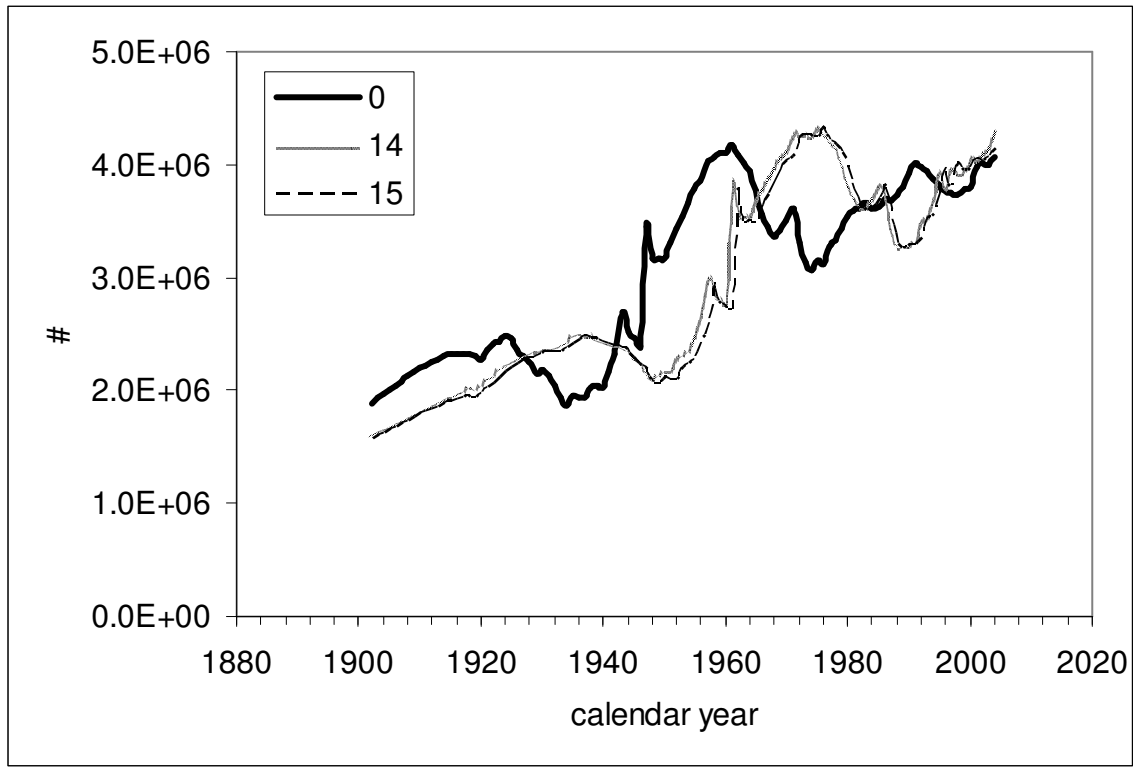

b)

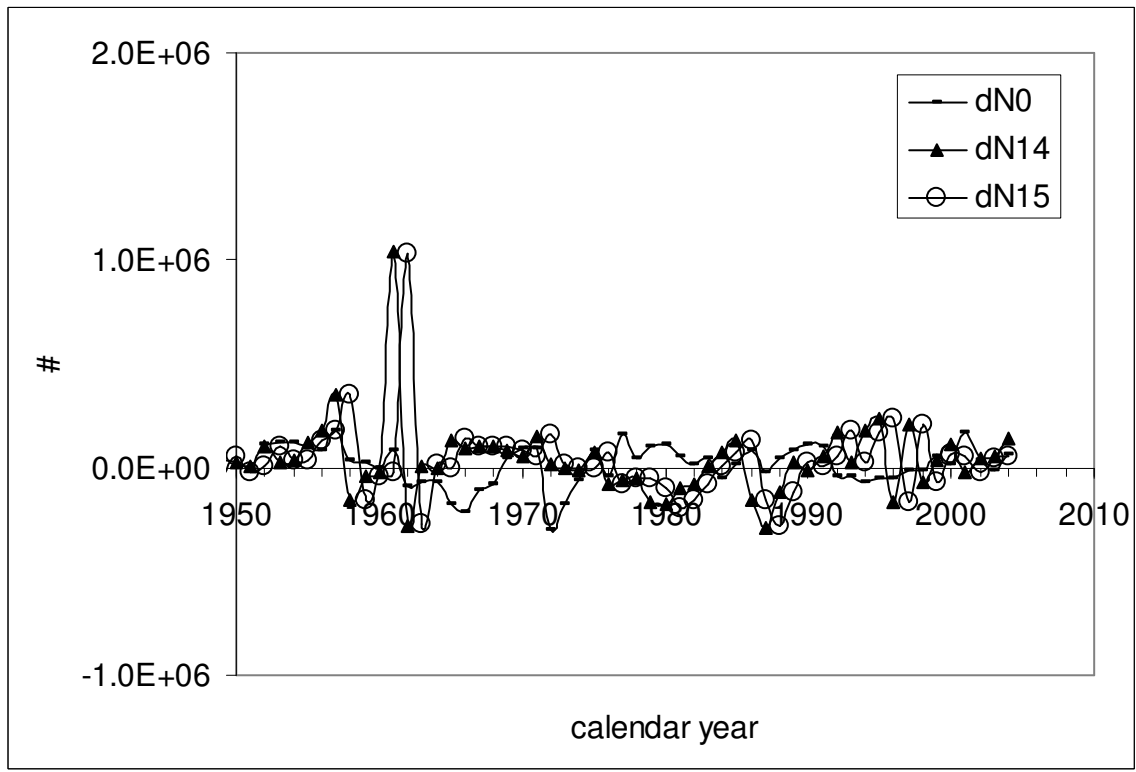

c)

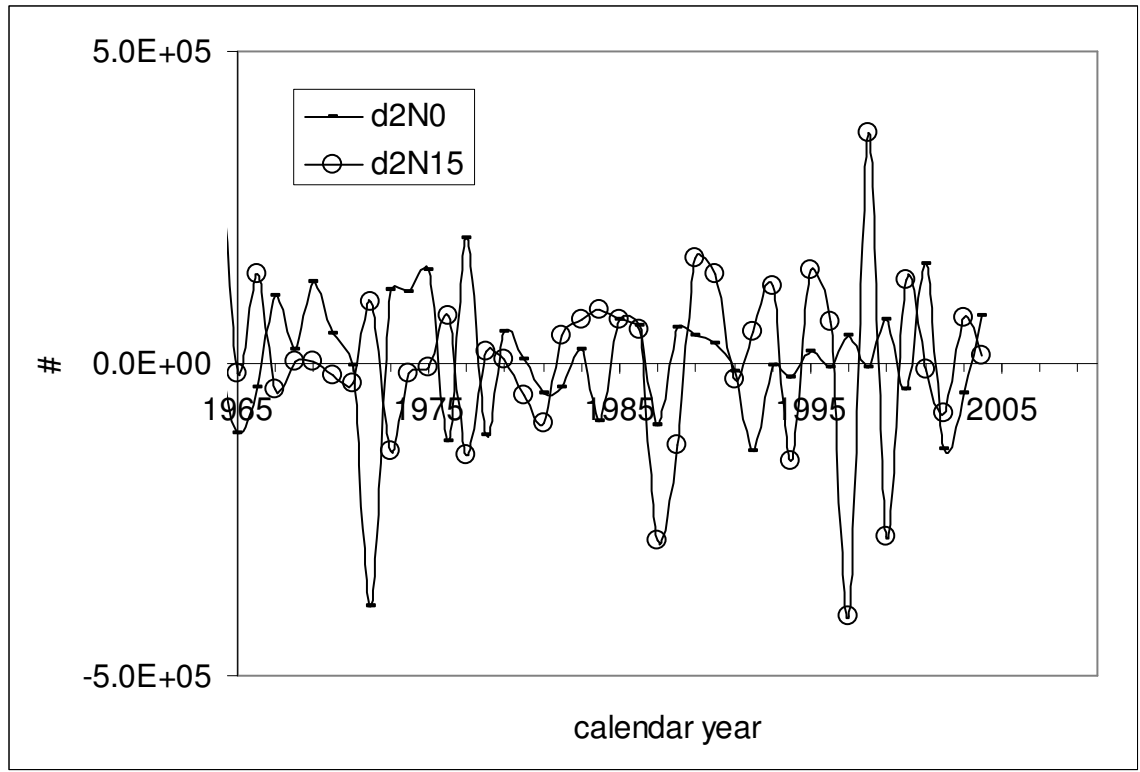

Figure A1.
a) The number of 15-, 14-, and 0-year-olds as a function of time between 1900 and 2004. Notice a large step in the earlier 1960s in the 15- and 14-year-olds corresponding to the burst in birth rate after the WWII.
b) The first differences of the time series presented in panel a). The step near 1960 is transformed into a spike, which is associated with natural causes. The spikes near the years of decennial censuses are of artificial character.
c) The second differentials of the series *N15* and *N0*.

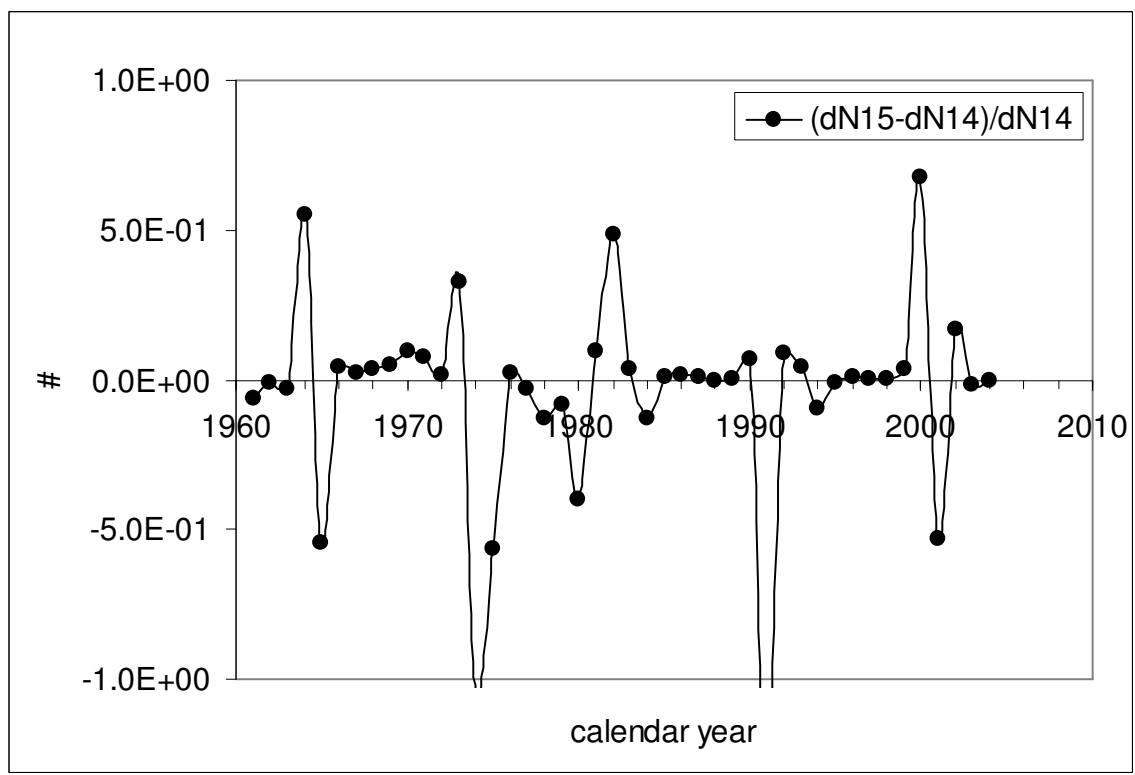

Figure A2. The ratio *[dN15(t)-dN14(t-1)]/dN14(t-1)* as a function of time. The years between decennial censuses are characterized by low (<0.1) values. The years near the censuses demonstrate artificial spikes induced by the CB procedures related to the correction of the age structure.

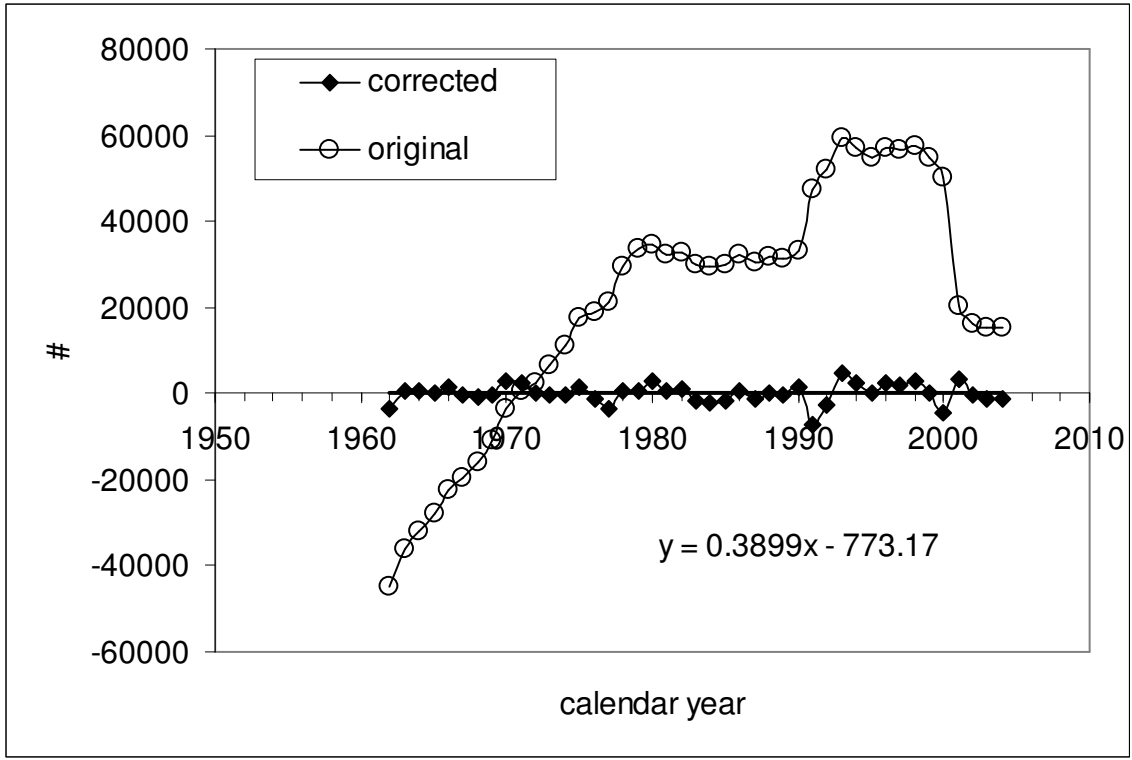

Figure A3. The original (open circles) and corrected (solid diamonds) difference between *N15(t)* and *N14(t-1)*. The latter is obtained by elimination of the linear trend and non-zero constants from the former. There are two low-amplitude spikes survived near 1990 and 2000, however.

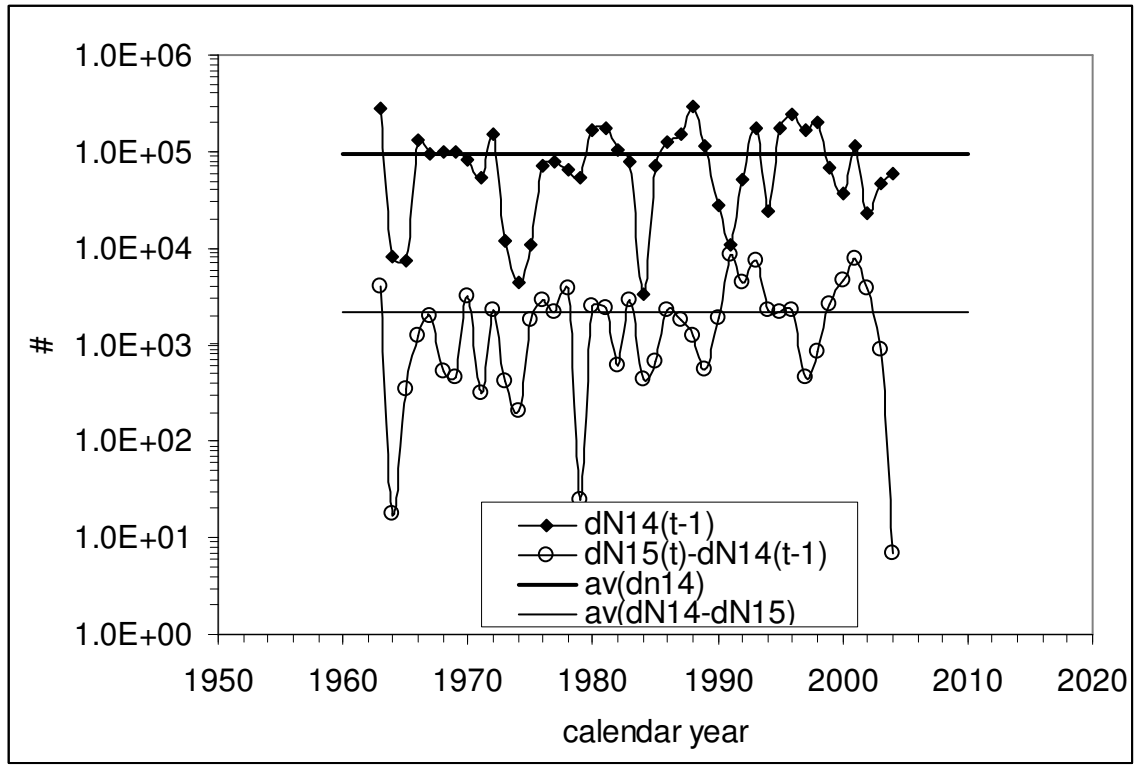

Figure A4. Comparison of abs*(dN15(t)-dN14(t-1))* and abs(*dN14*). The average values are 2200 and 95000, respectively. Their ratio is ~0.023.

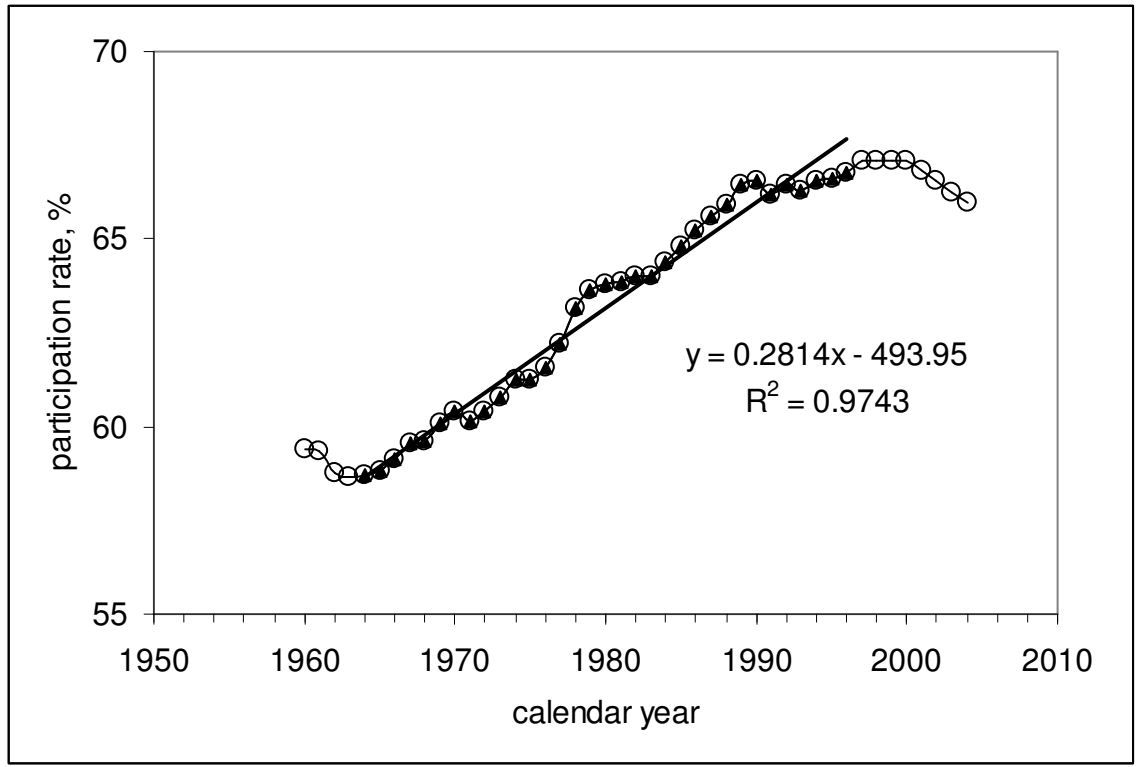

Figure A5. Labor force participation rate in the USA between 1960 and 2004. The years between 1965 and 1996 are characterized by a linear growth. Linear trend explains more than 97% of the participation rate variation during this period. In 2000, a period of slight decrease in the participation rate started.